\documentclass[runningheads]{llncs}

\usepackage{booktabs} % For formal tables
\usepackage{capt-of}

\usepackage{graphicx}
\usepackage[group-separator={,}]{siunitx}
\usepackage{mathtools}
\usepackage{floatflt}
\usepackage{paralist}
\usepackage{xint}
\usepackage{xintfrac}
\usepackage{xintexpr}
\usepackage{xspace}
\usepackage{float}
\usepackage{tikz}
\usepackage[nomessages]{fp}
\usepackage[hidelinks]{hyperref}

\usetikzlibrary{calc}

\newcommand{\classname}[1]{\ensuremath{\mathbf{#1}\xspace}}
\newcommand{\mathtext}[1]{\ensuremath{\mathrm{\text{#1}}}}

\newcommand{\blindtextRight}[1]{\makebox[0pt][r]{#1}}
\newcommand{\printWithLengthOf}[2]{\blindtextRight{#1}\phantom{#2}}

\newcommand{\set}[1]{\ensuremath{\left\{#1\right\}}}

\newcommand{\couplingMeasure}[4]{\ensuremath{\mathsf{coupdeg}_{#1,#2,#3}(#4)}}
\newcommand{\cptriple}[3]{\ensuremath{#1:#2\leftrightarrow #3}\xspace}

\extrafloats{100}

\newcommand{\resultFebTwentySeventeenpspcaller}{0.32782735902671073}

\newcommand{\resultFebTwentySeventeenpspcallee}{0.3007114441474247}

\newcommand{\resultFebTwentySeventeenpspboth}{0.29063440611252606}

\newcommand{\resultFebTwentySeventeenpsqcaller}{0.3561219979372329}

\newcommand{\resultFebTwentySeventeenpsqcallee}{0.31742406701888065}

\newcommand{\resultFebTwentySeventeenpsqboth}{0.3261118945883938}

\newcommand{\resultFebTwentySeventeenpdqcaller}{0.07957965859100381}

\newcommand{\resultFebTwentySeventeenpdqcallee}{0.2077606348270854}

\newcommand{\resultFebTwentySeventeenpdqboth}{0.22649392746637478}

\newcommand{\resultFebTwentySeventeencspcaller}{0.3064545960218466}

\newcommand{\resultFebTwentySeventeencspcallee}{0.409568851646706}

\newcommand{\resultFebTwentySeventeencspboth}{0.34650424357528636}

\newcommand{\resultFebTwentySeventeencsqcaller}{0.3648570683237015}

\newcommand{\resultFebTwentySeventeencsqcallee}{0.40607923732966}

\newcommand{\resultFebTwentySeventeencsqboth}{0.4069567305767922}

\newcommand{\resultFebTwentySeventeencdqcaller}{0.12981724124128546}

\newcommand{\resultFebTwentySeventeencdqcallee}{0.24396555174760215}

\newcommand{\resultFebTwentySeventeencdqboth}{0.288512395325373}

\newcommand{\resultSepTwentySeventeenpspcaller}{0.313960113960114}

\newcommand{\resultSepTwentySeventeenpspcallee}{0.2981810212579443}

\newcommand{\resultSepTwentySeventeenpspboth}{0.2823405654174885}

\newcommand{\resultSepTwentySeventeenpsqcaller}{0.34808678500986195}

\newcommand{\resultSepTwentySeventeenpsqcallee}{0.32887135656366423}

\newcommand{\resultSepTwentySeventeenpsqboth}{0.32773175542406313}

\newcommand{\resultSepTwentySeventeenpdqcaller}{0.08599605522682446}

\newcommand{\resultSepTwentySeventeenpdqcallee}{0.2176199868507561}

\newcommand{\resultSepTwentySeventeenpdqboth}{0.23442472057856673}

\newcommand{\resultSepTwentySeventeencspcaller}{0.3024068770873269}

\newcommand{\resultSepTwentySeventeencspcallee}{0.4123970726331556}

\newcommand{\resultSepTwentySeventeencspboth}{0.34424913143918523}

\newcommand{\resultSepTwentySeventeencsqcaller}{0.3630165123564579}

\newcommand{\resultSepTwentySeventeencsqcallee}{0.42689343670694513}

\newcommand{\resultSepTwentySeventeencsqboth}{0.412376635908102}

\newcommand{\resultSepTwentySeventeencdqcaller}{0.14421705354163272}

\newcommand{\resultSepTwentySeventeencdqcallee}{0.25815334787186434}

\newcommand{\resultSepTwentySeventeencdqboth}{0.3113053893454815}

\newcommand{\resultFebTwentyEighteenpspcaller}{0.3667177438937594}

\newcommand{\resultFebTwentyEighteenpspcallee}{0.2841875511121159}

\newcommand{\resultFebTwentyEighteenpspboth}{0.30199201185661534}

\newcommand{\resultFebTwentyEighteenpsqcaller}{0.39195212696179815}

\newcommand{\resultFebTwentyEighteenpsqcallee}{0.3140467956715539}

\newcommand{\resultFebTwentyEighteenpsqboth}{0.33116464451087274}

\newcommand{\resultFebTwentyEighteenpdqcaller}{0.06079864687214784}

\newcommand{\resultFebTwentyEighteenpdqcallee}{0.200998657478348}

\newcommand{\resultFebTwentyEighteenpdqboth}{0.234369112512246}

\newcommand{\resultFebTwentyEighteencspcaller}{0.375669305245959}

\newcommand{\resultFebTwentyEighteencspcallee}{0.37646133364179335}

\newcommand{\resultFebTwentyEighteencspboth}{0.35764880713350883}

\newcommand{\resultFebTwentyEighteencsqcaller}{0.4241123481503595}

\newcommand{\resultFebTwentyEighteencsqcallee}{0.3971352279043188}

\newcommand{\resultFebTwentyEighteencsqboth}{0.4040213625414009}

\newcommand{\resultFebTwentyEighteencdqcaller}{0.11672282760872248}

\newcommand{\resultFebTwentyEighteencdqcallee}{0.22310719272573104}

\newcommand{\resultFebTwentyEighteencdqboth}{0.2780108907271895}

\newcommand{\resultSepTwentyEighteenpspcaller}{0.36142762404292517}

\newcommand{\resultSepTwentyEighteenpspcallee}{0.281786941580756}

\newcommand{\resultSepTwentyEighteenpspboth}{0.29896907216494845}

\newcommand{\resultSepTwentyEighteenpsqcaller}{0.38910251873520646}

\newcommand{\resultSepTwentyEighteenpsqcallee}{0.31607256599589517}

\newcommand{\resultSepTwentyEighteenpsqboth}{0.3323713437186272}

\newcommand{\resultSepTwentyEighteenpdqcaller}{0.06232749062256718}

\newcommand{\resultSepTwentyEighteenpdqcallee}{0.2024791547073271}

\newcommand{\resultSepTwentyEighteenpdqboth}{0.235608818849754}

\newcommand{\resultSepTwentyEighteencspcaller}{0.3680415326477333}

\newcommand{\resultSepTwentyEighteencspcallee}{0.3784532412148486}

\newcommand{\resultSepTwentyEighteencspboth}{0.3499668558557259}

\newcommand{\resultSepTwentyEighteencsqcaller}{0.42142088315472154}

\newcommand{\resultSepTwentyEighteencsqcallee}{0.39958526248479515}

\newcommand{\resultSepTwentyEighteencsqboth}{0.4037933367949883}

\newcommand{\resultSepTwentyEighteencdqcaller}{0.12388112220122792}

\newcommand{\resultSepTwentyEighteencdqcallee}{0.22997588861163598}

\newcommand{\resultSepTwentyEighteencdqboth}{0.28781422331520773}

\newcommand{\drawPercentileChart}[7]{
\begin{center}
 \begin{figure}
 \fontsize{5}{1}\selectfont
 \begin{tikzpicture}
  \draw [->] (0cm,0cm) -- (7.75cm,0cm);
  \draw [->] (0cm,0cm) -- (0cm,3.5cm);
    -
   \foreach \basePercent/\baseXCoordinate/\number in {#2}
    {-
      \def\yScaleFactor{0.2}
      \def\xCoordinateBarLeftBorder{1.5mm + 0.075 * \baseXCoordinate cm}
      \def\xCoordinateBarRightBorder{2.5mm + 0.075 * \baseXCoordinate cm}
      \def\labelText{\number\%}
     \if\xintGeq{#1}{\number}1%
       \def\yCoordinate{\number}
       
       \def\yLabelOffset{0.175 cm}
     \else
       \def\yCoordinate{#1}
       \def\yLabelOffset{1.475 cm}
     \fi
     \fill[fill=blue]    (\xCoordinateBarLeftBorder, 0cm) rectangle (\xCoordinateBarRightBorder, \yScaleFactor * \yCoordinate cm);
     \if\xintGeq{#1}{\number}1%
     \else
       \fill[shade,top color=white,bottom color=blue]  (\xCoordinateBarLeftBorder, \yScaleFactor * \yCoordinate cm) rectangle (\xCoordinateBarRightBorder, 3.25cm);
     \fi
     \node[rotate=90] at (2.0mm + 0.075 * \baseXCoordinate cm, \yScaleFactor * \yCoordinate cm + \yLabelOffset)  {\color{gray} \labelText};
     \node[rotate=90] at (2.0mm + 0.075 * \baseXCoordinate cm, 0mm) {\printWithLengthOf{\basePercent\%}{aa}};
    };
    
    \node at (1cm,-7mm) {$n=#5$, $\mathtext{max}=#6$
    };
  \draw[dashed] (2mm + 0.075 * #7 cm, 0) -- (2mm + 0.075 * #7 cm, 4cm);

\end{tikzpicture} 
\caption{#3}#4
\end{figure}
\end{center}
}

\renewcommand{\drawPercentileChart}[7]{
\begin{figure}
Chart left out to speed up compilation.
\caption{#3}#4
\end{figure}
}

\renewcommand{\drawPercentileChart}[7]{}

\newcommand{\showExamplePoint}[5]{
\node[label={#4}] at (xyz cs:x=#1+0.5mm, y=#2-0.25mm, z=#3) { };

\draw[dashed,color=#5] (xyz cs:x=#1, y=#2, z=#3) -- (xyz cs:x=0, y=#2, z=#3);
\draw[dashed,color=#5] (xyz cs:x=#1, y=#2, z=#3) -- (xyz cs:x=#1, y=0,  z=#3);
\draw[dashed,color=#5] (xyz cs:x=#1, y=#2, z=#3) -- (xyz cs:x=#1, y=#2, z=0);

\draw[dashed,color=#5] (xyz cs:x=0, y=#2, z=#3) -- (xyz cs:x=0, y=0,  z=#3);
\draw[dashed,color=#5] (xyz cs:x=0, y=#2, z=#3) -- (xyz cs:x=0, y=#2, z=0);

\draw[dashed,color=#5] (xyz cs:x=#1, y=0, z=#3) -- (xyz cs:x=0, y=0, z=#3);
\draw[dashed,color=#5] (xyz cs:x=#1, y=0, z=#3) -- (xyz cs:x=#1, y=0, z=0);

\draw[dashed,color=#5] (xyz cs:x=#1, y=#2, z=0) -- (xyz cs:x=0, y=#2, z=0);
\draw[dashed,color=#5] (xyz cs:x=#1, y=#2, z=0) -- (xyz cs:x=#1, y=0, z=0);

\draw[dashed,color=#5] (xyz cs:x=0, y=0, z=0) -- (xyz cs:x=0, y=0,  z=#3);
\draw[dashed,color=#5] (xyz cs:x=0, y=0, z=0) -- (xyz cs:x=#1, y=0, z=0);
\draw[dashed,color=#5] (xyz cs:x=0, y=0, z=0) -- (xyz cs:x=0, y=#2, z=0);
}

\newcommand{\showAnalysisCoordinateSystem}{
\begin{tikzpicture}[x=0.5cm,y=0.5cm,z=0.3cm,>=stealth]
\sffamily
\draw[->] (xyz cs:x=0) -- (xyz cs:x=14.5) node[right] {$\alpha$: granularity};
\draw[->] (xyz cs:y=0) -- (xyz cs:y=14.5) node[above,align=left] {$\beta$: measurement approach};
\draw[->] (xyz cs:z=0) -- (xyz cs:z=23.5) node[above] {$\gamma$: coupling direction};

\foreach \coord\granularity in {6/\textbf class, 12/\textbf package}
{
  \draw (\coord,-1.5pt) -- (\coord,1.5pt) node[below] {\granularity};
}

\foreach \coord\granularity in {4/\textbf export, 8/\textbf import, 12/\textbf combined}
{
  \draw (xyz cs:y=-0.15pt,z=\coord) -- (xyz cs:y=0.15pt,z=\coord) node[below,rotate=45,pos=0.5] {\granularity};
}

\foreach \coord\approach in {4/\textbf static, 8/dynamic \\ \textbf unweighted, 12/dynamic \\ \textbf weighted}
{
 \draw (-1.5pt,\coord) -- (1.5pt,\coord) node[left,align=right] {\approach};
}

\showExamplePoint{12}{8}{8}{$(p,u,i$)}{gray}
\end{tikzpicture}
}

\newcommand{\avgCouplingFPEvalCall}[2]{(\csname result#1#2caller\endcsname+\csname result#1#2callee\endcsname+\csname result#1#2both\endcsname)/3}

\newcommand{\couplingKendallTauTableDataSetOne}{
\FPeval\cspString{\avgCouplingFPEvalCall{FebTwentySeventeen}{csp}}
\FPeval\csqString{\avgCouplingFPEvalCall{FebTwentySeventeen}{csq}}
\FPeval\cdqString{\avgCouplingFPEvalCall{FebTwentySeventeen}{cdq}}
\FPeval\pspString{\avgCouplingFPEvalCall{FebTwentySeventeen}{psp}}
\FPeval\psqString{\avgCouplingFPEvalCall{FebTwentySeventeen}{psq}}
\FPeval\pdqString{\avgCouplingFPEvalCall{FebTwentySeventeen}{pdq}}
\sisetup{round-mode=places}
\begin{small}
\begin{tabular}{l|S[round-precision=2]S[round-precision=2]S[round-precision=2]S[round-precision=2]S[round-precision=2]S[round-precision=2]}
          & \rotatebox{90}{\cptriple csu}         & \rotatebox{90}{\cptriple csw}        & \rotatebox{90}{\cptriple cuw}       & \rotatebox{90}{\cptriple psu}       & \rotatebox{90}{\cptriple psw}       & \rotatebox{90}{\cptriple puw}      \\ \hline
 import   & \resultFebTwentySeventeencspcaller    & \resultFebTwentySeventeencsqcaller   & \resultFebTwentySeventeencdqcaller  & \resultFebTwentySeventeenpspcaller  & \resultFebTwentySeventeenpsqcaller  & \resultFebTwentySeventeenpdqcaller \\
 export   & \resultFebTwentySeventeencspcallee    & \resultFebTwentySeventeencsqcallee   & \resultFebTwentySeventeencdqcallee  & \resultFebTwentySeventeenpspcallee  & \resultFebTwentySeventeenpsqcallee  & \resultFebTwentySeventeenpdqcallee \\
 combined & \resultFebTwentySeventeencspboth      & \resultFebTwentySeventeencsqboth     & \resultFebTwentySeventeencdqboth    & \resultFebTwentySeventeenpspboth    & \resultFebTwentySeventeenpsqboth    & \resultFebTwentySeventeenpdqboth   \\ 
 average  & \cspString                            & \csqString                           & \cdqString                          & \pspString                          & \psqString                          & \pdqString       
\end{tabular}
\end{small}

\captionof{table}{Coupling Analyses (Data Set 1)}
\label{tab:Coupling Analyses Feb Twentyseventeen}
}

\newcommand{\couplingKendallTauTableDataSetTwo}{
\FPeval\cspString{\avgCouplingFPEvalCall{SepTwentySeventeen}{csp}}
\FPeval\csqString{\avgCouplingFPEvalCall{SepTwentySeventeen}{csq}}
\FPeval\cdqString{\avgCouplingFPEvalCall{SepTwentySeventeen}{cdq}}
\FPeval\pspString{\avgCouplingFPEvalCall{SepTwentySeventeen}{psp}}
\FPeval\psqString{\avgCouplingFPEvalCall{SepTwentySeventeen}{psq}}
\FPeval\pdqString{\avgCouplingFPEvalCall{SepTwentySeventeen}{pdq}}
\sisetup{round-mode=places}
\begin{small}
\begin{tabular}{l|S[round-precision=2]S[round-precision=2]S[round-precision=2]S[round-precision=2]S[round-precision=2]S[round-precision=2]}
          & \rotatebox{90}{\cptriple csu}         & \rotatebox{90}{\cptriple csw}        & \rotatebox{90}{\cptriple cuw}       & \rotatebox{90}{\cptriple psu}       & \rotatebox{90}{\cptriple psw}       & \rotatebox{90}{\cptriple puw}      \\ \hline
 import   & \resultSepTwentySeventeencspcaller    & \resultSepTwentySeventeencsqcaller   & \resultSepTwentySeventeencdqcaller  & \resultSepTwentySeventeenpspcaller  & \resultSepTwentySeventeenpsqcaller  & \resultSepTwentySeventeenpdqcaller \\
 export   & \resultSepTwentySeventeencspcallee    & \resultSepTwentySeventeencsqcallee   & \resultSepTwentySeventeencdqcallee  & \resultSepTwentySeventeenpspcallee  & \resultSepTwentySeventeenpsqcallee  & \resultSepTwentySeventeenpdqcallee \\
 combined & \resultSepTwentySeventeencspboth      & \resultSepTwentySeventeencsqboth     & \resultSepTwentySeventeencdqboth    & \resultSepTwentySeventeenpspboth    & \resultSepTwentySeventeenpsqboth    & \resultSepTwentySeventeenpdqboth   \\ 
 average  & \cspString                            & \csqString                           & \cdqString                          & \pspString                          & \psqString                          & \pdqString       
\end{tabular}
\end{small}

\captionof{table}{Coupling Analyses (Data Set 2)}\label{tab:Coupling Analyses Sep Twentyseventeen}
}

\newcommand{\couplingKendallTauTableDataSetThree}{
\FPeval\cspString{\avgCouplingFPEvalCall{FebTwentyEighteen}{csp}}
\FPeval\csqString{\avgCouplingFPEvalCall{FebTwentyEighteen}{csq}}
\FPeval\cdqString{\avgCouplingFPEvalCall{FebTwentyEighteen}{cdq}}
\FPeval\pspString{\avgCouplingFPEvalCall{FebTwentyEighteen}{psp}}
\FPeval\psqString{\avgCouplingFPEvalCall{FebTwentyEighteen}{psq}}
\FPeval\pdqString{\avgCouplingFPEvalCall{FebTwentyEighteen}{pdq}}
\sisetup{round-mode=places}
\begin{small}
\begin{tabular}{l|S[round-precision=2]S[round-precision=2]S[round-precision=2]S[round-precision=2]S[round-precision=2]S[round-precision=2]}
          & \rotatebox{90}{\cptriple csu}         & \rotatebox{90}{\cptriple csw}        & \rotatebox{90}{\cptriple cuw}       & \rotatebox{90}{\cptriple psu}       & \rotatebox{90}{\cptriple psw}       & \rotatebox{90}{\cptriple puw}      \\ \hline
 import   & \resultFebTwentyEighteencspcaller    & \resultFebTwentyEighteencsqcaller   & \resultFebTwentyEighteencdqcaller  & \resultFebTwentyEighteenpspcaller  & \resultFebTwentyEighteenpsqcaller  & \resultFebTwentyEighteenpdqcaller \\
 export   & \resultFebTwentyEighteencspcallee    & \resultFebTwentyEighteencsqcallee   & \resultFebTwentyEighteencdqcallee  & \resultFebTwentyEighteenpspcallee  & \resultFebTwentyEighteenpsqcallee  & \resultFebTwentyEighteenpdqcallee \\
 combined & \resultFebTwentyEighteencspboth      & \resultFebTwentyEighteencsqboth     & \resultFebTwentyEighteencdqboth    & \resultFebTwentyEighteenpspboth    & \resultFebTwentyEighteenpsqboth    & \resultFebTwentyEighteenpdqboth   \\ 
 average  & \cspString                            & \csqString                           & \cdqString                          & \pspString                          & \psqString                          & \pdqString       
\end{tabular}
\end{small}

\captionof{table}{Coupling Analyses (Data Set 3)}\label{tab:Coupling Analyses Feb Twentyeighteen}
}

\newcommand{\couplingKendallTauTableDataSetFour}{
\FPeval\cspString{\avgCouplingFPEvalCall{SepTwentyEighteen}{csp}}
\FPeval\csqString{\avgCouplingFPEvalCall{SepTwentyEighteen}{csq}}
\FPeval\cdqString{\avgCouplingFPEvalCall{SepTwentyEighteen}{cdq}}
\FPeval\pspString{\avgCouplingFPEvalCall{SepTwentyEighteen}{psp}}
\FPeval\psqString{\avgCouplingFPEvalCall{SepTwentyEighteen}{psq}}
\FPeval\pdqString{\avgCouplingFPEvalCall{SepTwentyEighteen}{pdq}}
\sisetup{round-mode=places}
\begin{small}
\begin{tabular}{l|S[round-precision=2]S[round-precision=2]S[round-precision=2]S[round-precision=2]S[round-precision=2]S[round-precision=2]}
          & \rotatebox{90}{\cptriple csu}        & \rotatebox{90}{\cptriple csw}       & \rotatebox{90}{\cptriple cuw}      & \rotatebox{90}{\cptriple psu}      & \rotatebox{90}{\cptriple psw}      & \rotatebox{90}{\cptriple puw}     \\ \hline
 import   & \resultSepTwentyEighteencspcaller    & \resultSepTwentyEighteencsqcaller   & \resultSepTwentyEighteencdqcaller  & \resultSepTwentyEighteenpspcaller  & \resultSepTwentyEighteenpsqcaller  & \resultSepTwentyEighteenpdqcaller \\
 export   & \resultSepTwentyEighteencspcallee    & \resultSepTwentyEighteencsqcallee   & \resultSepTwentyEighteencdqcallee  & \resultSepTwentyEighteenpspcallee  & \resultSepTwentyEighteenpsqcallee  & \resultSepTwentyEighteenpdqcallee \\
 combined & \resultSepTwentyEighteencspboth      & \resultSepTwentyEighteencsqboth     & \resultSepTwentyEighteencdqboth    & \resultSepTwentyEighteenpspboth    & \resultSepTwentyEighteenpsqboth    & \resultSepTwentyEighteenpdqboth   \\ 
 average  & \cspString                           & \csqString                          & \cdqString                         & \pspString                         & \psqString                         & \pdqString       
\end{tabular}
\end{small}

\captionof{table}{Coupling Analyses (Data Set 4)}\label{tab:Coupling Analyses Sep Twentyeighteen}
}

\begin{document}
\title{Comparing Static and Dynamic Weighted Software Coupling Metrics}

\author{Henning Schnoor and Wilhelm Hasselbring}
\institute{Kiel University, Software Engineering Group \email{\{hs|wha\}@informatik.uni-kiel.de}}
\maketitle

\begin{abstract}
Coupling metrics are an established way to measure software architecture quality with respect to modularity. 
Static coupling metrics are obtained from the source or compiled code of a program, while dynamic metrics use runtime data gathered e.g., by monitoring a system in production. We study \emph{weighted} dynamic coupling that takes into account how often a connection is executed during a system's run. 
We investigate the correlation between dynamic weighted metrics and their static counterparts. 
We use data collected from four different experiments, each monitoring production use of a commercial software system over a period of four weeks. We observe an unexpected level of correlation between the static and the weighted dynamic case as well as revealing differences between class- and package-level analyses.
\keywords{software metrics, monitoring, dynamic analysis, static analysis}
\end{abstract}
\section{Introduction}

Coupling~\cite{chidamber1991towards,Stevens:1979:SD:1241515.1241533}---the number of inter-module connections in software systems---has long been identified as a software architecture quality metric for modularity \cite{Parnas1972}. Taking coupling metrics into account during development of a software system can help to increase the system's maintainability and understandability \cite{bogner2017automatically}, in particular for microservice architectures \cite{EMISA2019}. As a consequence, aiming for high cohesion and low coupling is accepted as a design guideline in software engineering \cite{Candela:2016:UCC:2943790.2928268}.

In the literature, there exists a wide range of different approaches to  measuring coupling. Usually, the coupling degree of a module (class or package) indicates the number of ``connections'' it has to different system modules. A ``connection'' between modules \classname A and \classname B can be, among others, a method call from \classname A to \classname B or an exception of type \classname B thrown by \classname A. Many notions of coupling can be measured statically, based on either source code or compiled code. 

Static analysis is attractive since it can be performed immediately on source code or on a compiled program. However, it has been observed~\cite{DBLP:journals/tse/ArisholmBF04,DBLP:journals/tse/CarverCN98,DBLP:journals/tse/ChidamberK94} that for object-oriented software, static analysis does not suffice, as it often fails to account for effects of inheritance with polymorphism and dynamic binding. This is addressed by dynamic analysis, where monitoring logs are generated while running the software. 

The results obtained by dynamic analysis depend on the workload used for the run of the system yielding the monitoring data. Hence the availability of representative workload for the system under test is crucial for dynamic analysis. As a consequence, dynamic analysis is more expensive than static analysis.

Dynamic analysis is often used to improve upon the accuracy of static coupling analysis \cite{Cornelissen+2009}. Dynamic analysis uses monitoring data to find, e.g., all classes \classname B whose methods are called by the class \classname A. In this case, the \emph{individual relationship} between two classes \classname A and \classname B is \emph{qualitative}: The analysis only determines whether there is a connection between \classname A and \classname B, and does not take its strength (e.g., number of calls during a system's run) into account. In contrast, a \emph{quantitative} coupling measurement quantifies the strength  of the connection between \classname A and \classname B by assigning it a concrete number. 

The coupling metrics we consider in this paper are defined using a \emph{dependency} graph. The nodes of such a graph are program modules (classes or packages). Edges between modules express call relationships. They can be labelled with \emph{weights}, which are integers denoting the number of occurrences of the call represented by the edge. Depending on whether coupling metrics take these weights into account or not, we call the metrics \emph{weighted} or \emph{unweighted}. The main two metrics we consider are the following:

\begin{enumerate}
 \item Unweighted static coupling, where an edge from \classname A to \classname B is present in the dependency graph if some method from \classname B is called from \classname A in the (source or compiled) program code,
 \item Weighted dynamic coupling, where an edge from \classname A to \classname B is present in the graph if such a call actually occurs during the monitored run of the system, and is attributed with the number of such calls observed.
\end{enumerate}

Dynamic weighted coupling measures cannot replace their static counterparts in their role to e.g., indicate maintainability of software projects. However, we expect dynamic weighted coupling measures to be highly relevant for software restructuring: In contrast to static coupling measures, weighted dynamic measures can reflect the runtime communication ``hot spots'' within a system, and therefore may be helpful in establishing performance predictions of restructuring steps. For example, method calls that happen infrequently may be replaced by a sequence of nested calls or with a network query without relevant performance impacts. Since static coupling measures are often used as basis for restructuring decisions~\cite{Candela:2016:UCC:2943790.2928268,DBLP:conf/icsm/MitchellM01}, dynamic weighted coupling measures can potentially complement their static counterparts in the restructuring process. This possible application leads to the following question: Do dynamic coupling measures yield additional information beyond what we can obtain from static analysis? 

Initially, we expected static and dynamic coupling degrees to be almost unrelated: A module \classname A has high static coupling degree if there are many method calls from \classname A to methods outside of \classname A or vice versa in the program code. On the other hand, \classname A has high dynamic weighted coupling degree if during the observed run of the system, there are many runtime method calls between \classname A and other parts of the system. Since a single occurrence of a method call in the code can be executed millions of times---or not at all---during a run of the program, static and weighted dynamic coupling degrees do not need to correlate.
Thus, our initial hypothesis was to not observe a high correlation between static and weighted dynamic metrics.

Our main research question is: \textbf{Are static coupling degrees and dynamic weighted coupling degrees statistically independent? If we observe correlation, can we quantify the correlation?}
 
To answer these questions, we compare the two coupling measures. We use dynamically collected data to compute weighted metrics that take into account the number of function calls during the system's run. We obtained the data from a series of four experiments. Each experiment consists of monitoring real production usage of a commercial software system (Atlassian Jira~\cite{Jira-730743771}) over a period of four weeks each. Our monitoring data contains more than three billion method calls. We compare the results from our dynamic analysis to computations of static coupling degrees. 

Directly comparing static and weighted dynamic coupling degrees is of little value, as these are fundamentally different measurements: For instance, the absolute value of dynamic weighted degrees depends on the duration of the monitored program run, which clearly is not the case for the static measures. We therefore instead compare \emph{coupling orders}, i.e., the ranking obtained by ordering all program modules by their coupling degree using the Kendall Tau\footnote{See~\cite{Andrew-KendallTauSpearman-EPS-1993} for a discussion of the relationship between this metric and Spearman's correlation.} metric~\cite{Kendall-Metric-1938}. This also allows to quantify the difference between such orders.

Our answer to the above stated research questions is that static and (weighted) dynamic coupling degrees are \emph{not} statistically independent. A possible interpretation of this result is that dynamic weighted coupling degrees give additional, but related information compared to the static case. In addition to this result, we observe insightful differences between class- and package-level analyses.

\subsection*{Contributions}

The results and contributions of this paper are:\footnote{A replication package inlcuding the collected data of our experiments will soon be published on Zenodo, to allow other researchers to repeat and extend our work.}
 \begin{itemize}
   \item Using a unified framework, we introduce precise definitions of static and dynamic coupling measures.
   \item To investigate our main research question, we performed four experiments involving real users of a commercial software product (the Atlassian Jira project and issue tracking tool \cite{Jira-730743771}) over a period of four weeks each. The software was instrumented via the dynamic monitoring framework Kieker \cite{KiekerICPE2012} based on AspectJ \cite{Kiczales2001}. From the collected data, we computed our dynamic coupling measures. We compared the obtained results, using the Kendall-Tau metric~\cite{Briand1996}, to coupling measures we obtained by static analysis.
   \item The results show that all coupling metrics we investigate are correlated, but there are also significant differences. In particular, when considering package-level coupling, the correlation is significantly stronger than for class-level coupling. As reason we assume that effects like polymorphism and dynamic binding often do not cross package boundaries.
\end{itemize}

Finally, we note that this paper is an extension of a previous short poster paper \cite{DBLP:conf/icse/SchnoorH18} in which a high-level overview of the research approach and the first data set are presented. The current paper extends the previous short poster paper (2 pages in length) as follows:

\begin{itemize}
 \item This paper contains an in-depth explanation of the research approach, including a precise definition of our coupling metrics.
 \item We report on the statistical properties of the data collected during the experiments.
 \item We report on the findings of four experiments whereas the short paper only discusses the first of our four data sets.
\end{itemize}

\subsection*{Paper Organization}

The remainder of the paper is organized as follows: In Section~\ref{sect:related work}, we discuss related work. Section~\ref{sect:dynamic, weighted metrics} provides our definition of weighted dynamic coupling. In Section~\ref{sect:static and dynamic analysis}, we explain our approach to static and dynamic analysis. Section~\ref{sect:jirakieker experiment summary} then describes the setting of our experiment. The results are presented and discussed in Section~\ref{sect:experiment results}. In Section~\ref{sect:threats to validity}, we discuss threats to validity and conclude in Section~\ref{sect:conclusion} with a discussion of possible future work.

\section{Related Work}\label{sect:related work}

There is extensive literature on using coupling metrics to analyse software quality, see, e.g., Fregnan et al.~\cite{DBLP:journals/infsof/FregnanBPB19} for an overview. Briand et al.~\cite{DBLP:journals/jss/BriandWDP00} propose a repeatable analysis procedure to investigate coupling relationships. Nagappan et al.~\cite{mining-metrics-to-predict-component-failures} show correlation between metrics and external code quality (failure prediction). They argue that no single metric provides enough information (see also Voas and Kuhn~\cite{DBLP:journals/computer/VoasK17}), but that for each project a specific set of metrics can be found that can then be used in this project to predict failures for new or changed classes. Misra et al.~\cite{DBLP:journals/iee/MisraAP12} propose a framework for the evaluation and validation of software complexity measures.
Briand and W\"ust~\cite{DBLP:journals/ac/BriandW02} study the relationship between software quality models to external qualities like reliability and maintainability. They conclude that, among others, import and export coupling appear to be useful predictors of fault-proneness.
Static weighted coupling measures have been considered by Offutt et al.~\cite{DBLP:journals/sqj/OffuttAS08}.
Allier et al.~\cite{DBLP:conf/scam/AllierVDS10} compare static and unweighted dynamic metrics.

Our approach is different: We do not study correlation between software metrics and software quality, but correlation between different software metrics.

Dynamic (unweighted) metrics have been investigated in numerous papers (see, e.g.,  Arisholm et al.~\cite{DBLP:journals/tse/ArisholmBF04} as a starting point, also the surveys by Chhabra and Gupta~\cite{DBLP:journals/jcst/ChhabraG10} and Geetika and Singh~\cite{Geetika:2014:DCM:2579281.2579296}). 
None of these approaches considers dynamic \emph{weighted} metrics, as we do.

Dynamic analysis is often used to complement static analysis. As an notable exception, Yacoub et al.~\cite{DBLP:conf/metrics/YacoubAR99} use weighted metrics. However, to obtain the data, they do not use runtime instrumentation---as we do---but ``early-stage executable models.'' They also assume a fixed number of objects during the software's runtime.

Arisholm et al.~\cite{DBLP:journals/tse/ArisholmBF04} study dynamic metrics for object-oriented software. Our dynamic coupling metrics are based on their \emph{dynamic messages} metric. The difference is as follows: Their metric counts only distinct messages, i.e., each method call is only counted once, even if it appears many times in a concrete run of the system. The main feature of our \emph{weighted} metrics is that the number of occurrences of each call during the run of a system is counted. The \emph{dynamic messages} metric from~\cite{DBLP:journals/tse/ArisholmBF04} corresponds to our unweighted dynamic coupling metrics (see below).

\section{Dynamic, Weighted Coupling}\label{sect:dynamic, weighted metrics}

\subsection{Dependency Graphs}\label{subsect:output of analyses}

We performed our analyses with two different levels of granularity: on the (Java) class and package levels. In the following we use the term \textit{module} for either class or package, depending on the granularity of the analysis. The output of either types of analyses (dynamic and static) is a labeled, directed graph $G$, where the nodes represent program modules (i.e., classes or packages), and the labels are integers which we refer to as \textit{weights} of the edges. An edge from $\classname A$ to $\classname B$ has label (weight) $n_{A,B}$, this denotes that the number of \textit{directed interactions} between $\classname A$ and $\classname B$ occurring in the analysis is $n_{A,B}$.

In the case of a static analysis, this means that there are $n_{A,B}$ places in the code of $\classname A$ where some method from $\classname B$ is called. For dynamic analysis, this means that during the monitored run of the system, there were $n_{A,B}$ run-time invocations of methods from $\classname B$ by methods from $\classname A$.

Our graph $G$ is a \textit{weighted dependency graph}, hence we call the coupling metrics we define below \textit{weighted} metrics. When we disregard the numbers $n_{A,B}$, the graph $G$ is a plain \textit{dependency graph}, i.e., a directed graph where the edges reflect function calls between the modules. We refer to metrics defined on the unweighted dependency graph---i.e., metrics that do not take the weights $n_{A.B}$ into account---as \textit{unweighted} metrics. We study the following three conceptually different approaches to measure coupling dependency between program modules:

\begin{enumerate}
 \item The first approach is \textit{static analysis}, which identifies method calls by analyzing the compiled code (we used BCEL to analyze Java \texttt{.class} and \texttt{.jar} files). Here we do not take weights into account. We therefore compute our static coupling measures from an unweighted dependency graph.
 \item Our second approach is \textit{unweighted dynamic analysis}. This analysis identifies method calls between modules as they appear in an actual run of the system (the data is obtained by monitoring), but does not take the weights $n_{A,B}$ into account. It therefore does not distinguish between cases where a module \classname A calls another module \classname B a million times or just once. This metric is essentially the  \emph{dynamic messages} metric from~\cite{DBLP:journals/tse/ArisholmBF04}.
 \item Our third approach is \textit{weighted dynamic analysis}, which differs from its unweighted counterpart only by taking the weights $n_{A,B}$ into account. 
\end{enumerate}

The distinctions between static/dynamic analyses and unweighted/weighted analyses are orthogonal choices. In particular, we omit in the present paper a weighted, static analysis, since our main motivation is the comparison of dynamic, weighted metrics unweighted, static metrics. 

\subsection{Definition of Coupling Metrics}\label{sect:measurement definitions}

We now define the coupling measures we study. Our measures assign a \textit{coupling degree} to a program module (i.e., a class or a package). We consider $18$ different ways to measure coupling, resulting from the following three orthogonal choices:

\begin{enumerate}
 \item The first choice is between \textbf{c}lass-level and \textbf{p}ackage-level granularity. Depending on this choice, a \textbf{module} is either a (Java) class or a (Java) package.
 \item The second choice is between one of our three basic measurement approaches: \textbf{s}tatic, dynamic \textbf{u}nweighted, or dynamic \textbf{w}eighted analysis.
 \item The third choice is to measure \textbf{i}mport- \textbf{e}xport- or \textbf{c}ombined coupling.
\end{enumerate}

\begin{figure}
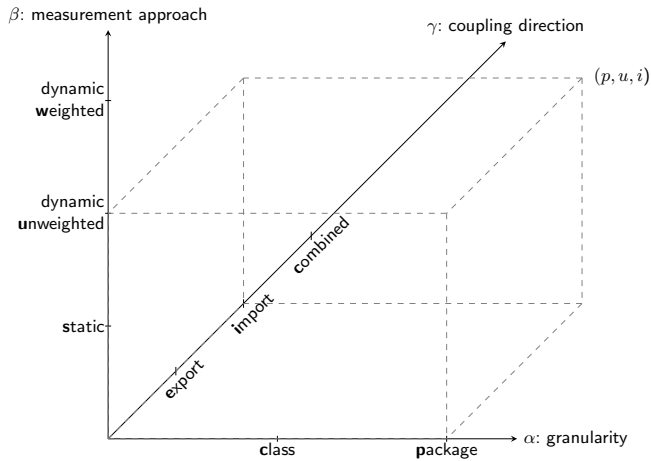
%{6cm}
\begin{center}
\scalebox{0.75}{\showAnalysisCoordinateSystem}
\caption{Dimensions of Analyses}
\label{fig:analysis coordinate system}
\end{center}
\end{figure}

To distinguish these 18 types of measurement, we use triples $(\alpha,\beta,\gamma)$, where $\alpha$ is \textbf{c} or \textbf{p} and indicates the granularity, $\beta$ is \textbf{s}, \textbf{u}, or \textbf{w} and indicates the basic measurement approach, and $\gamma$ is \textbf{i}, \textbf{e}, or \textbf{c}, indicating the direction of couplings taken into account. Figure~\ref{fig:analysis coordinate system} illustrates these three orthogonal choices: The example triple $(p,u,i)$ denotes an analysis with granularity \textbf package-level, using dynamic \textbf unweighted analysis, and considers coupling in the \textbf import direction.

Our coupling measures can be computed from the two dependency graphs resulting from our two analyses (static and dynamic). For a module $\classname A$, and a choice of measure $(\alpha,\beta,\gamma)$, the \textit{$(\alpha,\beta,\gamma)$-coupling degree of $\classname A$}, denoted with $\couplingMeasure{\alpha}{\beta}{\gamma}{\classname A}$, is computed as follows:

\begin{itemize}
 \item We compute $G_{\alpha,\beta}$. This is the weighted dependency graph between classes (if $\alpha=c$) or packages (if $\alpha=p$) obtained by static analysis (if $\beta=s$) or dynamic analysis (if $\beta=u$ or $\beta=w$), where each weight $n_{A,B}$ is replaced with $1$ if the analysis is static or (dynamic) unweighted (i.e., if $\beta\in\set{s,u}$).
 \item Then, $\couplingMeasure{\alpha}{\beta}{\gamma}{\classname A}$ is the out-degree of $\classname A$, in-degree of $\classname A$, or sum of these, depending on whether $\gamma=i$, $\gamma=e$, or $\gamma=c$.
 The in (out) degree of $\classname A$ is the sum of the weights of its incoming (outgoing) edges in the graph. 
\end{itemize}

\section{Static and Dynamic Analysis}\label{sect:static and dynamic analysis}

We perform our static analysis (using the Apache BCEL~\cite{BCEL-60}) on the compiled code. This also implies that some optimizations have already performed by the compiler, such as removal of dead code. Therefore, our static and dynamic analyses are performed on the exact same code, without differences introduced in the compilation process. For the dynamic analysis, we use the Kieker framework~\cite{KiekerICPE2012} that allows to register every method call. Kieker uses AspectJ's \cite{Kiczales2001} load-time weaver to instrument the analyzed software automatically at load-time. In order to reduce the performance impact of monitoring, we restricted the monitoring to a subset of the system, and adjusted the static analysis accordingly.

\section{Experiment Design}\label{sect:jirakieker experiment summary}

We analyzed the software Atlassian Jira, versions 7.3.0, 7.4.3, and 7.7.1~\cite{Jira-730743771}. The system was instrumented using AspectJ technology. For each method call, we recorded the time stamp, the class name of caller and of the callee.

To perform our analysis with realistic workload, we conducted four experiments with real users using a software system (Atlassian Jira \cite{Jira-730743771}) in production. Jira was used by students participating in a mandatory programming course of our computer science curriculum. In the course, the students develop a software using the Kanban process management method~\cite{Kanban2013}. The time span of the project is four weeks, with full time participation by the students. 

\begin{floatingtable}{
\begin{tabular}{lllr}
 \textbf{\#} & \textbf{date}   & \textbf{users} & \textbf{method calls} \\ \hline
           1 & February 2017   & 19             & \num{196442044} \\
           2 & September 2017  & 48             & \num{854657027} \\
           3 & February 2018   & 16             & \num{475357185} \\
           4 & September 2018  & 58             & \num{2409688701} \\
\end{tabular}}
\caption{Number of users and monitored method calls}
\label{tab:number of method calls}
\end{floatingtable}

We report on four experiment runs, from February and September of 2017 and 2018. Each time, the software ran for a four-week period. The collected monitoring data from each run includes the startup sequence, basic configuration such as database access, initial tasks as user registration and setup of the Kanban boards, and day-to-day usage. No person-related data is used for our analysis. In Table~\ref{tab:number of method calls}, we list the number of method calls recorded as well as the number of users of our Jira installation in each of the three experiment runs. 

Obviously, there are differences between the four runs of the software that we analyze. For example, different students took parts in the course each time, the focus of the project required using different features of the Jira software in each iteration, and we also instructed them to use more features of the tool in the later iterations (this is one reason why the number of method calls per student is higher in the later runs of the experiment). Therefore, our four experiments---even though they are conducted using the same software system---give us slightly more variation in the data than running the exact same software with the exact same group of users. However, our main analysis results do not vary significantly between the different runs of the experiment, indicating that our findings are invariant under small changes of the experiment setup.

\section{Experiment Results}\label{sect:experiment results}

\subsubsection{Compared Measures}

We compare the coupling degrees computed by these different approaches. Comparing the actual ``raw'' values of $\couplingMeasure{\alpha}{\beta}{\gamma}{\classname A}$ for different combinations of $\alpha$, $\beta$, $\gamma$ and some class or package \classname A does not make much sense: The weighted values depend on the length of the measurement run of the system, whereas the static analysis does not.

However, the absolute coupling values are usually not the most interesting results of such an analysis. For a developer, the identification of the modules with the highest coupling degree is among the most interesting results of applying a software metric. Therefore, a useful approach is to study the relationship between the \emph{orders} among the modules in the different analyses: Each analysis yields an ordering of the classes or packages from the ones with the highest coupling degree to the ones with the lowest one; we call these orders \emph{coupling orders}. These orders can be compared between different analyses of varying measurement durations.

Given our coupling measure definitions, we have the following choices for a left-hand-side (LHS) and a right-hand-side (RHS) analysis:
 \begin{itemize}
 \item The first choice is whether to consider class or package analyses (both the LHS and the RHS should consider the same type of module).
 \item The second choice is which two of our three basic measurement approaches (see Section~\ref{subsect:output of analyses}) we intend to compare: \textit{s}tatic analysis, (dynamic) \textit{u}nweighted analysis, and (dynamic) \textit{w}eighted analysis. There are three possible choices: \textit{s} vs.~\textit{u}, \textit{s} vs.~\textit{w}, and \textit{u} vs.~\textit{w}.
 \item For each combination, we consider import, export, and combined coupling. 
\end{itemize}

Hence, there are $18$ comparisons we can perform in each of our four data sets, leading to $72$ different comparisons.

\subsubsection{Kendall-Tau distance}

To study the difference between our different basic measurement approaches, we compare the coupling orders of the analyses using the Kendall-Tau distance~\cite{Briand1996}: For a finite base set $S$ with size $n$, the metric compares two linear orders $<_1$ and $<_2$. The Kendall-Tau distance $\tau(<_1,<_2)$ is the number of swaps needed to obtain the order $<_1$ from $<_2$, normalized by dividing by number of possible swaps $\frac{n(n-1)}2$. Hence $\tau(<_1,<_2)$ is always between $0$ (if $<_1$ and $<_2$ are identical) and $1$ (if $<_1$ is ``reverse'' of $<_2$). Values smaller than $0.5$ indicate that the orders are closer together than expected from two random orders, while values larger than $0.5$ indicate the opposite. 

\subsubsection{Distance Values}\label{sect:kendall tau results}\label{sect:inter-analysis results}

To present our results, we use the following notation to specify the LHS and RHS analyses: We use a triple $\cptriple{\alpha}{\beta_1}{\beta_2}$, where
\begin{inparaitem}
 \item $\alpha$ is $c$ or $p$ expressing \textbf class or \textbf package coupling,
 \item $\beta_1$ is $s$ or $u$ expressing whether the LHS analysis is \textbf static or (dynamic) \textbf unweighted,
 \item $\beta_2$ is $u$ or $w$ expressing whether the RHS analysis is (dynamic) \textbf unweighted or (dynamic) \textbf weighted.
\end{inparaitem}

For each of these combinations, we consider export, import, and combined coupling analyses. This results in $18$ comparisons for each data set, which are presented in Tables~\ref{tab:Coupling Analyses Feb Twentyseventeen}, \ref{tab:Coupling Analyses Sep Twentyseventeen}, \ref{tab:Coupling Analyses Feb Twentyeighteen}, and \ref{tab:Coupling Analyses Sep Twentyeighteen} for our four experiments. 

\subsubsection{Statistical Significance}

To measure statistical significance, we computed the absolute z-scores of our experiments. The smallest observed absolute z-score among all our experiments is $9.41$, and all but two absolute values are above 10. As a point of reference, the corresponding likelihood for z-score 10 is $7.6\cdot10^{-24}$, this is the probability to observe the amount of correlation seen in our dataset under the assumption that the compared orders are in fact independent. This indicates a huge degree of statistical significance, which is due to the large number of program units appearing in our analysis.

\noindent
\scalebox{0.75}{
\begin{minipage}{7cm}
\couplingKendallTauTableDataSetOne 
\end{minipage}}
\scalebox{0.75}{
\begin{minipage}{7cm}
\couplingKendallTauTableDataSetTwo 
\end{minipage}
}

\ \\

\noindent
\scalebox{0.75}{
\begin{minipage}{7cm}
\couplingKendallTauTableDataSetThree 
\end{minipage}}
\scalebox{0.75}{
\begin{minipage}{7cm}
\couplingKendallTauTableDataSetFour 
\end{minipage}}

\subsubsection{Discussion}\label{sect:comparison results discussion}

The first obvious take-away from the values presented in Tables \ref{tab:Coupling Analyses Feb Twentyseventeen}-\ref{tab:Coupling Analyses Sep Twentyeighteen} is that all $72$ reported distances (and of course also the average values) are below $0.5$, many of them significantly so. This indicates that there is a significant similarity between the coupling orders of the static and the two dynamic analyses. This was not to be expected: While in small runs of a system, one could possibly conjecture that there might not be a large difference between the static and dynamic notions of coupling, this changes when we analyze longer system runs: In our longest experiment, we analyzed more than 2.4 billion method calls. The dynamic, weighted coupling degree of a class \classname{A} is the number of calls from or to methods from \classname{A} among these 2.4 billion calls, while its static, unweighted coupling degree is the number of classes \classname{B} such that the compiled code of the software contains a call from \classname{A} to \classname{B} or vice versa. A single method call in the code is only counted once in an unweighted analysis, but this call can be executed millions of times during the experiment, and each of these executions is counted in the weighted, dynamic coupling analysis. Therefore, it was not necessarily to be expected that we observe correlation between unweighted static and weighted dynamic coupling degrees.

However, our results suggest that all of the three types of analyses that we performed are correlated, with different degrees of significance. In particular, dynamic weighted coupling degrees seem to give additional, but not unrelated information compared to the static case.

The static coupling order is closer to the dynamic unweighted than to the dynamic weighted order in almost all cases. This was expected: In an hypothetical ``complete run'' of a system, and in the absence of 
issues resulting from object-oriented features these measures would coincide. On the other hand, the dynamic weighted analysis is very different from the static one by design. 

A very interesting observation is that in all $36$ cases except for $3$ cases involving import coupling in our first two data sets, comparing $\cptriple{c}{\beta}{\gamma}$ for some coupling measure with $\beta\in\set{\mathtext{\textbf{s},\textbf{u},\textbf{w}}}$ and $\gamma\in\set{\mathtext{\textbf{i},\textbf{e},\textbf{c}}}$.

to $\cptriple{p}{\beta}{\gamma}$ shows that the distance from the analysis of the package case is smaller than the corresponding distance in the class case, sometimes significantly so. A possible explanation is that in the package case, the object-oriented effects that are often cited as the main reasons for performing dynamic analysis are less present, as, e.g., inheritence relationships are often between classes residing in the same package.

\section{Threats to Validity}\label{sect:threats to validity}

Concerning external validity, our analysis is limited by the fact that we covered only four runs, each with four weeks, of only one software system (Atlassian Jira). 
To address this threat, we plan to monitor additional software tools such as Jenkins and Tomcat (which are also used in the course).
Concerning internal validity, our dynamic analysis omits some of Jira's classes in order to maintain sufficient performance of the system. To ensure that our comparisons in Section~\ref{sect:inter-analysis results} are conclusive, we only considered the classes and packages covered by both the static and dynamic analysis in the computation of the Kendall Tau distances. Additionally, different interpretations of what is considered as coupling between the static and in the dynamic analyses are always possible. However, since our notion of coupling is rather simple (method calls between different classes), we are confident that our static and dynamic analysis in fact use the same notion of coupling. Finally, as discussed in Section~\ref{sect:static and dynamic analysis}, we examine compiled code, not source code. When performing a similar analysis on source code, the differences between the static and the dynamic analyses would likely increase, as the dynamic analysis of course also uses compiled code. However, this can also be seen as an advantage, since this allows us to focus on the differences between static code and a running system, which is the goal of this study.

\section{Conclusions and Future Work}\label{sect:conclusion}

We studied three different basic measurement approaches: Static coupling, unweighted dynamic coupling, and weighted dynamic coupling. We performed four runs of an experiment that allows to compare these metrics to static coupling measurements. Our results, as discussed in Section~\ref{sect:comparison results discussion}, suggest that dynamic coupling metrics complement their static counterparts: Despite the large (and expected) difference, there is also a statistically significant correlation. This suggests that further study of dynamic weighted coupling and its relationship with other coupling metrics is an interesting line of research. 

A key question is how the additional information given by weighted dynamic coupling measurements can be used to evaluate the architectural quality of software systems, or more generally, to assist a software engineer in her design decisions. Coupling metrics can be used as recommenders for restructuring \cite{Candela:2016:UCC:2943790.2928268}, and for static coupling measures, correlation between coupling and external quality has been observed \cite{SMR:SMR1842}. A study of the relationship between static coupling measures and changeability and code comprehension has been performed in~\cite{DBLP:conf/metrics/YacoubAR99}. In~\cite{anuradha2015dynamic}, it is argued that unweighted dynamic metrics can be used for maintenance prediction. Since dynamic weighted metrics contain additional information compared to their unweighted counterparts, it will be interesting to study whether and how this additional information can be used in these contexts.

\end{document}